\documentclass[notitlepage,reprint,pra,nofootinbib,twocolumn]{revtex4-2} 
\usepackage{amsmath,amsthm,amssymb}
\usepackage{physics}
\usepackage{graphicx} 
\usepackage{mathtools}
\usepackage{epstopdf}

\usepackage{float}
\DeclareGraphicsExtensions{.png,.pdf}

\usepackage{hyperref}

 \usepackage{fancyhdr}
\begin{document}

\fancyhead[R]{\ifnum\value{page}<2\relax\else\thepage\fi}

\title[Article Title]{Resolution of 100 photons and quantum generation of unbiased random numbers}











\newcommand*{\UVA}{Department of Physics, University of Virginia, 382 McCormick Rd, Charlottesville, 22904-4714, VA, USA}
\newcommand*{\Jlab}{Thomas Jefferson National Accelerator Facility, Newport News, Virginia 23606, USA}    
\newcommand*{\NAS}{ National Academy of Sciences, 500 Fifth St. N.W., Washington D.C., 20001, USA}
\newcommand*{\AFRL}{Information Directorate, Air Force Research Laboratory, Rome, 13441, NY, USA}   
\newcommand*{\Lehman}{Department of Physics and Astronomy, Lehman College, The City University of New York, Bronx, 10468-1589, NY, USA}

\author{Miller Eaton$^1$}
\email{me3nq@virginia.edu}

\author{Amr Hossameldin$^1$}
\email{ah6sr@virginia.edu}
\thanks{* and $\dagger$ contributed equally to this work.}
\author{Richard J. Birrittella$^{2,5}$}
\author{Paul M. Alsing$^2$}
\author{Christopher C. Gerry$^3$}
\author{Hai Dong$^4$}
\author{Chris Cuevas$^4$}
\author{Olivier Pfister$^1$}
\address{$^1$\UVA}
\address{$^2$\AFRL}
\address{$^3$\NAS}
\address{$^4$\Lehman}
\address{$^5$\Jlab}

\begin{abstract}
Macroscopic quantum phenomena, such as observed in superfluids and superconductors, have led to promising technological advancements and some of the most important tests of fundamental physics. At present, quantum detection of light is mostly relegated to the microscale, where avalanche photodiodes are very sensitive to distinguishing single-photon events from vacuum but cannot differentiate between larger photon-number events. Beyond this, the ability to perform measurements to resolve photon numbers is highly desirable for a variety of quantum information applications including computation, sensing, and cryptography. True photon-number resolving detectors do exist, but they are currently limited to the ability to resolve on the order of 10 photons, which is too small for several quantum state generation methods based on heralded detection. In this work, we extend photon measurement into the mesoscopic regime by implementing a detection scheme based on multiplexing highly quantum-efficient transition-edge sensors to accurately resolve photon numbers between zero and 100. We then demonstrate the use of our system by implementing a quantum random number generator with no inherent bias. This method is based on sampling a coherent state in the photon-number basis and is robust against environmental noise, phase and amplitude fluctuations in the laser, loss and detector inefficiency as well as eavesdropping. Beyond true random number generation, our detection scheme serves as a means to implement quantum measurement and engineering techniques valuable for photonic quantum information processing.
\end{abstract}





\maketitle


\section{Main}\label{sec1}
The nature of quantum mechanics dictates a fundamental wave-particle duality for physical systems, which was first recognized by Einstein through the understanding that light is composed of individual energy quanta known as photons~\cite{Einstein1905_2}. The ability to accurately measure photons has led to checking the validity of the notion of `spooky action at a distance'~\cite{Salart2008} and tremendous technological advancement in quantum communication~\cite{Gisin2007}, quantum metrology~\cite{becerra2013experimental,Slussarenko2017,Nehra2019}, and quantum computation~\cite{Zhong2020,Arrazola2021}. Much of this progress relies on the ability to measure single photons, such as through the use of avalanche photodiodes (APDs)\cite{campbell2016recent}; however, the ability to resolve arbitrary numbers of photons beyond simply distinguishing vacuum from non-vacuum is highly desirable for many quantum information applications~\cite{Becerra2015,Arrazola2019,Thekkadath2020,Arrazola2021}. The process of projecting a subset of modes of an entangled state onto the Fock-basis can allow for engineering non-Gaussian quantum states with negative Wigner functions~\cite{Eaton2019,Ra2020,Walschaers2021} --- a requirement for any quantum speed-up in continuous-variable quantum information~\cite{Mari2012}. Recent claims of quantum supremacy with Gaussian boson sampling devices~\cite{Zhong2020} can be challenged with substantially greater ease when threshold detectors are used in place of photon-number-resolving detectors (PNRDs)~\cite{bulmer2021boundary}. Finally, sampling photon-number of a wave-like superposition such as a coherent state reveals fundamentally random outcomes that can be used to generate true random numbers~\cite{furst2010high,ren2011quantum,Gerry2022}. 

\begin{figure*}[ht]
    \includegraphics[width = \textwidth]{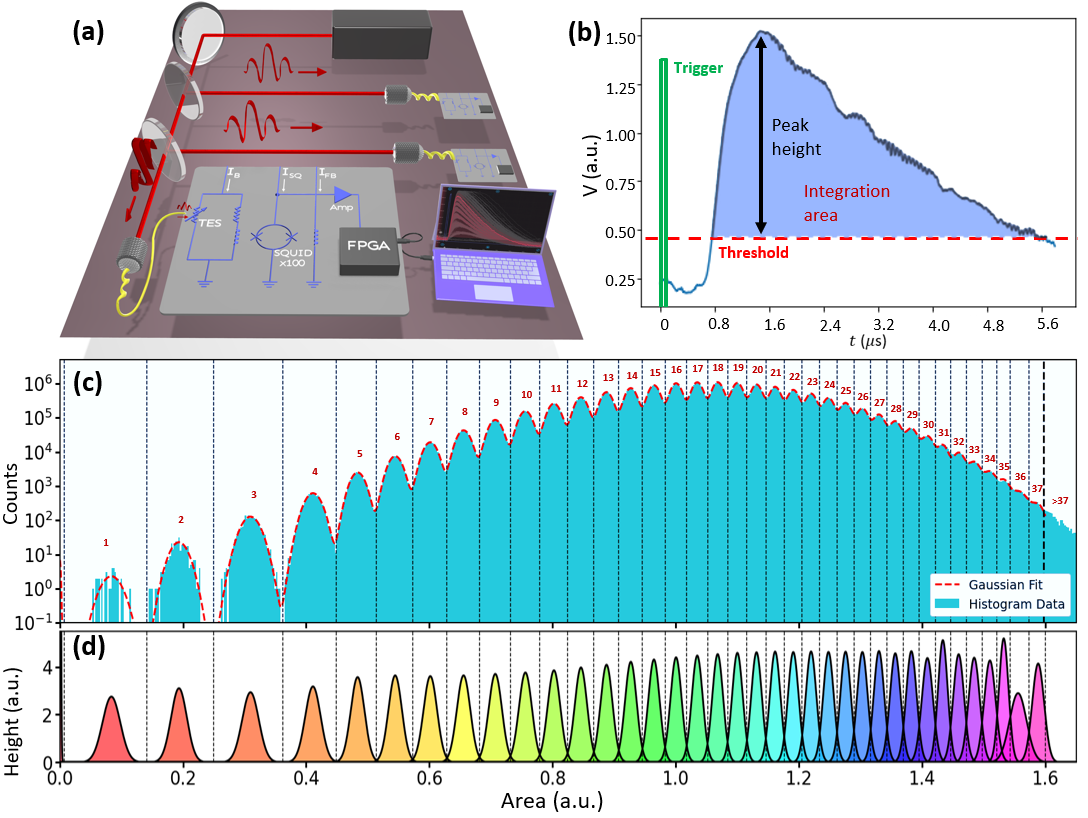}
\caption{(a) Experimental setup. A pulsed sources is evenly split into three segments and each is coupled to a TES detector channel. (b) Example event (blue) following the pulse trigger (green). Pulse parameters including area and height are recorded if the signal passes a specified threshold. (c) Histogram of measured signal areas of $10^8$ events for a single TES channel where a sum of Gaussians (dashed red line) is used to fit the data to determine binning for photon-number resolution. Bins are set at the intersection of between the normalized Gaussians as shown in (d). }
    \label{fig:exp}
\end{figure*}

The transition-edge sensor (TES), which is based on a calorimeter formed from a superconducting wafer held just below the critical temperature, has arisen as a viable PNRD with quantum efficiency approaching unity and entirely negligible dark counts~\cite{Lita2008,Fukuda2011,Gerrits2016}. 
Previous results with TES systems show the ability to measure non-classical systems with high mean-photon numbers~\cite{Gerrits2012,Harder2016}; however, these experiments were based on methods requiring extensive post-processing that give generally good estimates of photon-number measurements but relatively low distinguishability between individual photon counts above 10 photons~\cite{Levine2012}. For demanding applications requiring photon-number resolution, even a single photon discrepancy destroys quantum correlations. Current methods demonstrate the potential to accurately count photons in the low double-digits($\sim16$)~\cite{Morais2020}, but certain proposals necessitate considerably higher detection events for conditional state preparation.  One particularly salient example is the preparation of a cubic-phase state to complete a universal gate set for continuous-variable quantum computation~\cite{Gottesman2001}. In order for the numerical approximations used in this formalism to hold, one must detect a large number of photons --- simulations suggest 50 or more~\cite{Ghose2007}. The detection scheme we demonstrate here now easily surpasses this previously unreachable milestone.

In this work, we extend the resolving capabilities of individual TES detectors to a maximum of 37 photons per detection channel with on-the-fly signal processing. We then multiplex three detectors into a system capable of resolving 0-100 photons with detector quantum efficiencies above $90\%$. Furthermore, we illustrate the utility of our scheme toward quantum cryptography applications by creating a quantum random number generator (QRNG). The need for random numbers arises in many applications including cryptography, simulation, and games of chance. Pseudo-random number generators (PRNG) are not truly random and can, for example, lead to erroneous results in Monte Carlo simulations~\cite{ferrenberg1992monte}. The stochastic nature of quantum mechanics leads to true randomness, but many current implementations sample random events from a non-uniform distribution which can lead to bias that must be corrected classically~\cite{Ma2016,herrero2017quantum}. Our method to implement a QRNG is based on sampling the photon statistics of a coherent state and is fundamentally unbiased, robust to experimental and environmental noise, and invulnerable to eavesdropping.

The detection system used here is constructed by splitting a laser pulse equally across three paths and sending each to a TES as shown in Fig.~\ref{fig:exp}(a). Each TES is a PNRD that makes use of the extremely temperature-dependent resistance of a superconductor near the phase transition. Our TESs are comprised of superconducting tungsten wafers that operate with a critical temperature near 100 mK. When light is incident on a chip, the thermal energy of an absorbed photon acts to locally break the superconducting state and induce a spot of non-zero resistance, which increases nearly linearly with absorbed energy~\cite{Lita2008}. This change in resistance is detected by a series of highly sensitive superconducting quantum interference devices (SQuIDs) and is then amplified and converted to an output voltage that is sent to an external field-programmable gate array (FPGA) to extract key signal parameters on the fly (system details in Methods). The detectors used were optimized to be highly absorptive at the desired wavelength, and while our detectors achieve above $90\%$ quantum efficiency at the target wavelength of 1064 nm (details in Methods), TES systems have achieved efficiencies of $\eta=0.98$~\cite{Fukuda2011} and show the potential to reach $\eta>0.99$~\cite{fujii2012thin}.

\section{True photon-number resolving measurements}\label{sec2}

In order to resolve absorbed photon number, information to distinguish different outputs must be extracted from the signal received by the FPGA. An example signal is depicted in Fig.~\ref{fig:exp}(b). Traditionally, peak height has been used for an indicator as the magnitude of the voltage is proportional to the energy absorbed for low-photon numbers~\cite{Gerrits2016}. However, this technique limits individual detector resolution due to the saturation of the peak magnitudes beyond several photons, so recently, alternative methods have been explored for extracting useful information~\cite{Morais2020}. Although the maximum voltage of the peak saturates, the electrical resistance of the TES continues to change as it re-cools back to the superconducting state, suggesting useful information is contained beyond the peak as the cooling time will also depend on the energy absorbed. Integrating the signal in the region above a pre-defined noise threshold yields information about both the maximum voltage and the time to cool the TES; this peak area thus allows the resolution of many more photons than height alone. 

For a single TES channel, the histogram of areas for $10^8$ measurement events of a pulsed coherent state is shown in Fig.~\ref{fig:exp}(c). As the pulse area monotonically increases with absorbed energy, the distinctly separated bins correspond exactly to the quanta of energy detected and can be used to inform the number of photons measured. The location of these bins can be determined by fitting the obtained histogram to a sum of Gaussian functions (red dotted line in the figure), where the intersection of each normalized Gaussian gives the location of the bin edge. The reason for a Gaussian distribution within each bin is due to variations in the peak areas resulting from electronic and thermal noise on the cooling tail of signal peaks. The Gaussian fitting breaks down for large areas beyond the black dashed line in Fig.~\ref{fig:exp}(c) indicating the photon-number can no longer be accurately determined for this detector. The number of events beyond the detector resolution across all three TES channels accounts for less than $0.3\%$ of events.

The normalized Gaussian fits to the histogram are displayed in the bottom panel, Fig.~\ref{fig:exp}(d), where it can be seen that the overlap of neighboring Gaussian peaks is quite small for the majority of bins, indicating high confidence in correctly determining the true photon number for a given area measurement. The confidence rate decreases with photon number but remains above $90\%$ for photon numbers from 0-20 in Fig.~\ref{fig:exp}(d). If one is willing to post-select and slightly reduced count rates, the accuracy of a given photon-number assignment can be substantially increased by defining regions of uncertainty near the bin edges. If an event area is recorded in this uncertainty region, then the event is discarded and not considered in the statistics. Provided the regions of uncertainty are scaled in terms of the fitted Gaussian widths, $\sigma_n$, then the measured probability distribution will not deviate from the true distribution and the accuracy of individual photon-number assignment will increase. If the regions of uncertainty are defined beyond $\pm\sigma_n$, then $32\%$ of the data is discarded, but the confidence rates increase to $99\%$ or higher for the first 20 photons. If area events are only kept within $\pm \tfrac{1}{2}\sigma_n$ of each peak, then confidence rates further increase to $99\%$ out to 31 photons. The area histograms, Gaussian fits, and quantitative overlap errors for each of the three detection channels are given in the Extended Data.

Post-selection of data was not necessary for the QRNG experiment performed in this work as the results only required random parity measurements as will be described in the next section. Fortunately, the well-centered Gaussian distributions in each histogram bin mean that the probability to improperly count an $n$ photon event as an $n+1$ event is approximately the same as the probability to mistake an $n+1$ event for an $n$ photon count for all events away from the edge of the detector range. Due to this effect and the predominance of detection events away from the upper edge of the TES range, the statistical error for the QRNG experiment was dominated by finite sampling.

\section{Quantum random number generation}\label{sec3}

The prototypical photonic QRNG is based on sending a single photon to a balanced beamsplitter and placing detectors on the output to determine whether the photon was transmitted or reflected~\cite{stefanov2000optical,jennewein2000fast}. This is a truly random coin-flip in the ideal case, but it comes with limitations, such as the need for on-demand single photons, a perfectly balanced beamsplitter, and ideal detectors. Other optical techniques, such as homodyne measurements to detector random vacuum fluctuations~\cite{gabriel2010generator} or a variation on the first method where weak light is spread across a sensor array~\cite{sanguinetti2014quantum} can also be used, but these methods also suffer from physical limitations and noise that lead to randomness with bias. The randomness achieved is not sampled from a uniform distribution and therefore systematic bias must be removed with classical algorithms~\cite{vonNeumann1951,Peres1992rand}. Beyond reducing data and requiring vulnerable classical schemes, systems with inherent bias are at risk to quantum hacking~\cite{zhao2008quantum}, where an adversary can effectively change the calibrated bias and use this to their advanced to break encryption. 

\begin{figure*}[ht]
    \centering
    \includegraphics[width = \textwidth]{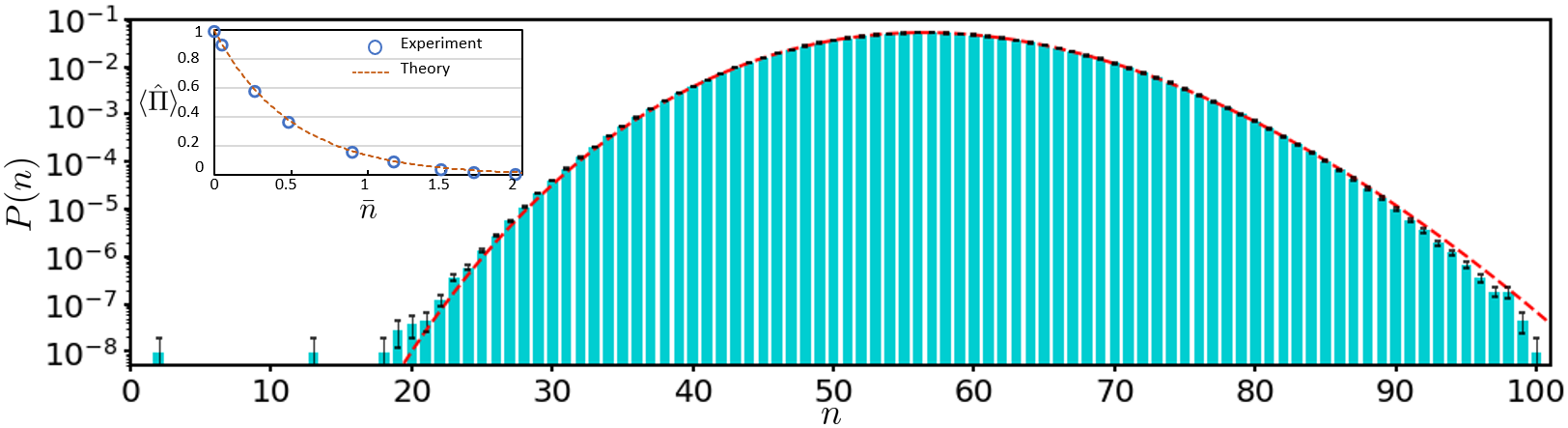}
\caption{Experimental photon number distribution obtained by splitting a coherent state of mean photon number $\Bar{n}= 57$ across three TES channels over $10^8$ events. The red dashed line indicates the theoretical Poissonian distribution with a mean of 57. Error bars shown are of one SD and are obtained from finite sampling and photon-number binning errors. The plot inset shows the measured parity coherent states begins near one (vacuum) but tends to zero as the amplitude increases. The measured parity for the $\Bar{n}=57$ coheret state is $\langle\hat{\Pi}\rangle=-7\times10^{-5}\pm 10^{-4}$.}
    \label{fig:results}
\end{figure*}

Here, we implement a QRNG making use of the inherent randomness present in the parity of the Poissonian distribution of a coherent state~\cite{ren2011quantum,Gerry2022}. When sampling the parity of the photon-number distribution, the inherent bias vanishes exponentially quickly with increasing coherent state intensity and asymptotically approaches a true coin flip. To generate the random numbers we simply convert a photon number detection to a binary output, where each even photon-number event is assigned an outcome of `0' and odd photon-numbers are assigned a `1'. This method is unaffected by experimental imperfections such as photon loss, detector inefficiency, phase and amplitude noise, and contamination by environmental noise.

For the parity operator given by $\hat{\Pi}=(-1)^{\hat{n}}=e^{i\pi\hat{n}}$ where $\hat{n}=\hat{a}^\dag\hat{a}$ is the photon-number operator and the operators $\hat{a}^\dag$ and $\hat{a}$ are the respective bosonic creation and annihilation operators, we can examine the expectation value of parity for a coherent state,
\begin{equation}
    \ket{\alpha} = e^{-\frac{1}{2}\abs{\alpha}^2} \sum_{n=0}^{\infty}\frac{\alpha^n}{\sqrt{n!}}\ket{n}.
\end{equation}
If $\Bar{n}=\langle\hat{n}\rangle$ is the mean photon number of the coherent state, then the expectation of parity is given by
\begin{equation}
    \langle \hat{\Pi}\rangle= P_e-P_o=e^{-2\Bar{n}},
\end{equation}
where $P_e$ and $P_o$ are the respective probabilities to detect either even or odd photon numbers.

In Fig.~\ref{fig:results}, we show the experimentally measured probability distribution for a large coherent state with $\Bar{n}=57$, which allows us to make full use of our PNRD and clearly resolve out to 100 photons. Although the theoretical parity of this state is $e^{-114}\sim10^{-50}$, we cannot hope to reach this precision due to finite sampling. With $10^8$ measurement events, we achieve a parity of zero to within uncertainty, with the measured value of $-7\times10^{-5}\pm 10^{-4}$. Additionally, we first verify the parity of weaker coherent states as shown in the inset of Fig.~\ref{fig:results}. As expected, the parity of vacuum is 1, and we are clearly able to match the trend of $e^{-2\Bar{n}}$ for increasing $\Bar{n}$.

One unfortunate downside of TES detection systems is the slow detector response leading to lower generation rates. Recent advances show that superconducting nanowire single-photon detectors (SNSPDs) have the potential to be used as PNRDs that are orders of magnitude faster than TESs~\cite{cahall2017multi}, but until this technology matures, we implement an alternative method to increasing random bit generation rates. As opposed to binning the photon number result by parity, a uniformly random distribution can also be obtained by taking the measurement result and binning according to photon-number modulo $2^d$ where $d\in\mathbb{Z}$. In this way, we can generate a bit string of size $d$ for each measurement. As $d$ increases, the residual bias of the QRNG still asymptotes to zero with increasing $\Bar{n}$, but a larger coherent state amplitude is needed to achieve a similarly negligible bias. In this work with a maximum detection of 100 photons, we find that the residual bias for a coherent state with $\Bar{n}=57$ is equivalent for $d\in\{1,2,3\}$, so we use modulo 8 binning to generate random numbers. 

\begin{figure*}
    \centering
    \includegraphics[width = 0.8\textwidth]{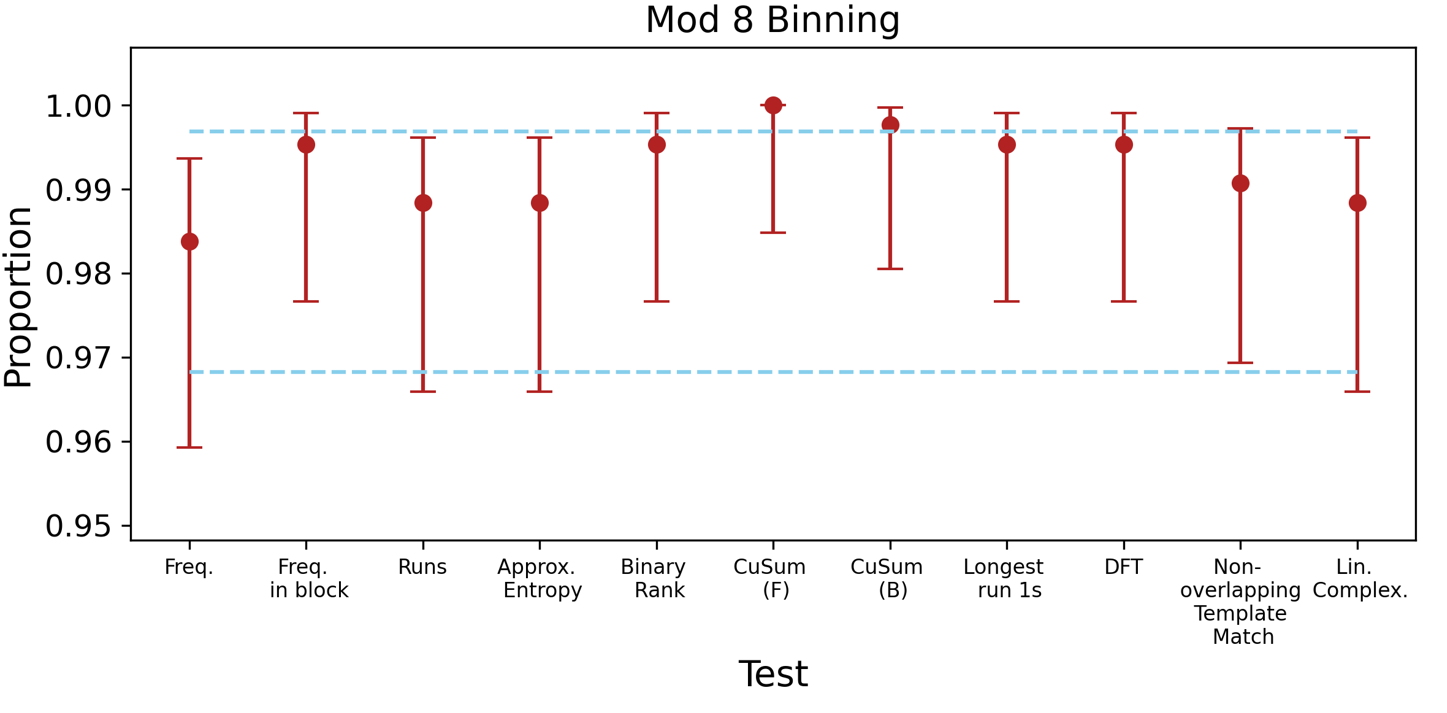}
\caption{Randomness tests for the resultant bit strings from $10^8$ events based on assigning three bits of information to each event by taking the measured photon number modulo 8. Data was broken into segments of $7.5\times10^{5}$ bits and each string was tested for randomness. The proportion (red markers), i.e. the percentage of trials that pass a test given a significance level of $\alpha$=0.01, falls within the corresponding confidence interval for all tests considered indicating evidence of true randomness. The error bars for each proportion are computed from the Wilson score (confidence) interval of Eq.~\ref{eqn:WI_interval} where $n=431$ is the total number of trials and $n_s\;\left(n_f\right)$ are the number of successful (failed) trials for a significance level of $\alpha=0.01$. Given repeated testing of the bit generation method, the error bars denote the probability range for which the proportion is likely to fall.}
    \label{fig:mod8randomness}
\end{figure*}

 We subject the $\sim3\times 10^8$ random bits generated by our protocol to a series of tests taken from the NIST suite of randomness tests. The proportion (i.e. the percentage of tests that pass a given test) is plotted in Fig.~\ref{fig:mod8randomness} for each test, given a significance level of $\alpha=0.01$. In computing the confidence interval for Fig.~\ref{fig:mod8randomness} (dashed blue lines), we do not make the standard approximation that the distribution of error about the binomially-weighted observation is given by that of a normal distribution, since our sample size is small enough that such an approximation will be unreliable. Instead we use the Wilson score (confidence) interval \cite{ref:Wilson}, which has been shown to be reliable for smaller sample sizes.  The findings in Fig.~\ref{fig:mod8randomness} demonstrate that our measurements indicate randomness across all tests considered (all proportions lie above the lower confidence bound). We additionally show the results of randomness measures for binning with $d\in[1,5]$ in the Extended Data.

\subsection{Robust nature of proposed method}

Upon closer examination we can see how our method here proves to be quite robust against various sources of error. First, we can consider phase and amplitude fluctuations originating either from the laser or any other experimental instability. This can be modeled by assuming that a statistical mixture of coherent states impinges upon the detector. We find that phase fluctuations have absolutely no bearing on the randomness and still lead to the same residual bias of $e^{-2\Bar{n}}$, which we experimentally verify as shown in the Extended Data. Amplitude fluctuations similarly provide negligible impact. Suppose the coherent state has mean photon number of $\Bar{n}$ and there is a small intensity fluctuation of $\delta$. The expectation of parity becomes $e^{-2(\Bar{n}\pm\delta)}\approx e^{-2\Bar{n}}(1\pm\delta)$ which tends to zero for sufficiently large $\Bar{n}$.

Next, we can consider the effects of loss, detector inefficiency, and uneven splitting between the TES channels with imperfect beamsplitters. We can always model an inefficient detector by inserting a loss channel in the form of a beamsplitter of transmittivity $\eta$ before a perfect detector and performing a partial trace over the unmeasured output port (Methods). As the coherent state, $\ket{\alpha}$, maps to the smaller coherent state, $\ket{\sqrt{\eta}\alpha}$, after this loss, an imperfect detector still measures a Poissonian photon-number distribution. Thus, in order to achieve quality randomness with low residual bias, the coherent state used must be chosen such that $\Bar{n}'=\eta\Bar{n}$ is sufficiently large. As for uneven splitting or differing detector efficiencies between channels, we can equivalently model the process of measuring a single coherent state distribution as the discrete convolution of three smaller coherent state distributions. As all beamsplitter outputs are still detected, changing the beamsplitter reflectivities just acts to redistribute the photons amongst the TES channels. Provided no single channel saturates, which is easily recognizable through monitoring area measurements, sampling the summed output of all channels will still yield a Poissonian distribution.

An additional concern of any quantum mechanical experiment is that of unintentional coupling to the environment. One possible effect of such coupling is photon loss as addressed in the previous paragraph. Another effect is the addition of photons, such as coupling to an external thermal bath, or some malicious observer attempting to inject light. In place of measuring a coherent state, suppose that the detector is sent the density operator $\rho = \rho_{\alpha} \otimes \rho_{env}$, where $\rho_{\alpha}=\ket{\alpha}\bra{\alpha}$ is the density operator for the coherent state and $\rho_{env}$ is the density operator for some unknown quantum state, not necessarily pure, originating from the environment. The expectation value of parity for the whole system is given by $\langle e^{i\pi \sum \hat{n}_k}\rangle$, where subscript $k$ denotes the different subsystems. This leads to an overall parity of 
\begin{equation}
    \langle\hat{\Pi}\rangle =e^{-2\Bar{n}}\langle \hat{\Pi} \rangle_{env},
\end{equation}
where $\langle \hat{\Pi} \rangle_{env}$ is the parity of the environment alone and is bounded between 1 and -1. Thus environmental mixing will not degrade the quality of the QRNG.

As a final concern, consider an eavesdropper attempting to determine information about the random numbers. Suppose an eavesdropper uses a beamsplitter to sample the coherent light in an attempt to predict the random number measured by the user. Due to the nature of coherent states, the two beamsplitter outputs remain in a product state, hence are not correlated. Thus no information about the results at one output port can be used to determine the results at the other, preventing the eavesdropper from attaining useful information. Other side-channel attacks, such as the insertion of different quantum states by a nefarious party, can be readily mitigated as well. Although the QRNG method only utilizes higher order parity measurements, we still have access to the full photon-number distribution from the TES, which can be monitored to ensure that Poissonian statistics are still obtained. This rules out any external manipulation since replacing or interspersing the coherent state with a different state will yield a different distribution. Additionally, the TES waveform response can be concurrently monitored and frequently recalibrated to rule out signal manipulation. Finally, as a coherent state is simply a laser output, the source and detector can be fabricated in near proximity to one another and protected from any realistic attack through appropriate shielding.

Recently, there has been some emphasis on the use of Bell inequality violations to certify the quantum nature of a device and ensure private randomness~\cite{Ma2016,Acin2016,herrero2017quantum}. Although this concept has merit, it requires closing all experimental loopholes to eliminate a local hidden variable theory before it can truly validate a black box as a quantum device. Furthermore, trust must be given at some point during any realistic experiment as the classical signal used to enact Bell measurements may themselves be spoofed. In our implementation, the quantum nature of the experiment is verified by the area histograms shown in Fig.~\ref{fig:exp}(c). The origin of the separation between area measurements is the fundamental energy quantization of photons. An entirely classical signal would yield a single broad Gaussian peak centered about the average energy of the beam of light spanning a swath of areas due to classical noise fluctuations as opposed to the multiple Gaussian fits for each TES channel.

In this Article, we have demonstrated drastic improvement to the photon-number resolving capabilities of high quantum efficiency TES systems and can accurately resolve 0-100 photons. By post-selecting data, one can achieve error rates below $1\%$ on photon-number measurements beyond 30 photons per detection channel without impacting the measurement distribution. These results have far-reaching implications for quantum information applications by opening up avenues in quantum sensing, such as reaching the Heisenberg limit with large photon-number parity detection~\cite{gerry2000heisenberg}, or through uses in photonic quantum computation, such as efficiently simulating interactions in quantum field theory~\cite{Marshall2015a}. Furthermore, we demonstrated the utility of our detection scheme to make an unbiased QRNG by sampling the parity of a coherent state. This technique is robust to a variety of experimental imperfections, and bit generation rates can be improved through binning with photon-number modulo $2^d$. 

\section{Acknowledgements}
M. E., A. H., and O. P. were supported by National Science Foundation grants No. DMR-1839175 and No. PHY-1820882.  M. E., A. H., C. C., H. D., and O. P. were additionally supported by Jefferson Lab LDRD project No. LDRD21-17 under which Jefferson Science Associates, LLC, manages and operates Jefferson Lab. R. J. B. acknowledges support from the National Research Council Research Associate Program. C. C. G. acknowledges support under the AFRL Summer Faculty Fellowship Program (SFFP). P. M. A. and C. C. G. acknowledge support from the Air Force Office of Scientific Research (AFOSR). M.E. thanks L. A. Morais and R. Nehra for discussion, and T. Gerrits for advice regarding the TES. O. P. thanks A. Miller, A. Lita, and S. W. Nam for building the initial single-channel detector system. Any opinions, findings, conclusions, or recommendations expressed in this material are those of the author(s) and do not necessarily reflect the views
of the Air Force Research Laboratory (AFRL). 
The appearance of external hyperlinks does not constitute endorsement by the United States Department of Defense or General Electric of the linked websites, or the information, products, or services contained therein.  The Department of Defense does not exercise any editorial, security, or other control over the information you may find at these locations.

\section{Author Contributions}
M.E., A.H., and O.P. designed the experimental setup and characterized the detector. H.D. and C.C. built and programmed the EFADC for data collection. M.E. and A.H. collected and analyzed measured data. R.J.B., P.M.A., and C.C.G. devised the method to make the unbiased QRNG. R.J.B. and P.M.A. performed the data analysis for characterizing randomness of the generated bit sequence. The article was written by M.E. with contributions from all authors. 

\section{Competing Interests}
The authors declare no competing interests.

\section{Methods}
\subsection{Theoretical background}
\subsubsection{Origin of randomness}
The photon-number parity of a coherent state tends towards a uniform distribution as the energy of the state increases. For a coherent state given by $\ket{\alpha} = e^{-\frac{1}{2}\abs{\alpha}^2} \sum_{n=0}^{\infty}\frac{\alpha^n}{\sqrt{n!}}\ket{n}$ and parity operator given by  $\hat{\Pi}=(-1)^{\hat{n}}=e^{i\pi\hat{n}}$ where $\hat{n}=\hat{a}^\dag\hat{a}$ is the photon-number operator. We can derive 
\begin{align*}
    \bra{\alpha}\hat{\Pi} \ket{\alpha} &=\bra{\alpha}e^{i\pi\hat{n}} \ket{\alpha} \\ \nonumber
    &=e^{-\abs{\alpha}^2} \sum_{n,n'=0}^{\infty} \frac{{\alpha^*}^{n'} \alpha^n}{\sqrt{n'!n!}} \bra{n'}e^{i\pi\hat{n}} \ket{n} \\ \nonumber
    &=e^{-\abs{\alpha}^2} \sum_{n,n'=0}^{\infty} \frac{{\alpha^*}^{n'} \alpha^n}{\sqrt{n'!n!}} e^{i\pi n} \bra{n'} \ket{n} \\ \nonumber
    &=e^{-\abs{\alpha}^2} \sum_{n=0}^{\infty} \frac{(\abs{\alpha}^2 e^{i\pi})^n}{n!} \\ \nonumber
    &=e^{-2\Bar{n}} 
\end{align*}
where $\Bar{n}=\bra{\alpha}\hat{n}\ket{\alpha}=\abs{\alpha}^2$.\\

From this we see that for large $\Bar{n}$, the parity expectation value can be arbitrarily close to zero. To generate the random numbers we simply output `0' whenever we measure an even number or `1' whenever we measure odd.
\subsubsection{Phase and amplitude fluctuations}
First, we consider phase fluctuations.  Suppose we do not have a pure coherent state, but a statistical mixture of coherent states with the same amplitude and a random phase, 
\begin{equation}
    \rho_{coh} = \frac{1}{2\pi} \int_0 ^{2\pi} d\phi \ket{r e^{i\phi}}\bra{r e^{i\phi}},
    \label{eq:rho_mix}
\end{equation}
where $r = \lvert\alpha\rvert = \sqrt{\Bar{n}}$. \\

This yields 
\begin{align*}
    \langle\hat{\Pi}\rangle &= \text{Tr}[\rho_{coh}\hat{\Pi}] \\ \nonumber
    &=\frac{1}{2 \pi}\int_{0}^{2 \pi} d\phi \sum_{n=0}^{\infty} \bra{n}\ket{r e^{i\phi}}\bra{r e^{i\phi}}e^{i\pi\hat{n}}\ket{n} \\ \nonumber
    &=\frac{1}{2 \pi}\int_{0}^{2 \pi} d\phi \sum_{n=0}^{\infty} e^{i\pi n} \abs{\bra{n} \ket{r e^{i\phi}}}^2 \\ \nonumber
    &=\frac{1}{2 \pi}\int_{0}^{2 \pi} d\phi \sum_{n=0}^{\infty} (-1)^n \abs{e^{-\frac{1}{2}\Bar{n}} \sum_{i=0}^{\infty}\frac{(\sqrt{\Bar{n}e^{i\phi}})^i}{\sqrt{i!}}}^2 \\ \nonumber 
    &=\frac{1}{2 \pi}\int_{0}^{2 \pi} d\phi \sum_{n=0}^{\infty} (-1)^n e^{-\Bar{n}} \frac{\Bar{n}^n}{n!}\\ \nonumber 
    &= e^{-2\Bar{n}}
\end{align*}

\noindent which shows that phase noise does not affect the parity expectation value. \\

Second, we consider amplitude fluctuations. Changes in the amplitude of the coherent state amount to changes in the mean photon number $\Bar{n}$. For a change $\delta$ in the mean photon number, the parity expectation value becomes $e^{-2(\Bar{n}\pm\delta)}$ which is approximately $e^{-2\Bar{n}}$ for small $\delta$.\\

\subsubsection{Environmental noise}
We now look at the expectation value of the parity operator on the whole system where $\rho = \rho_{coh} \otimes \rho_{env}$ with $\rho_{coh} = \ket{\alpha}\bra{\alpha}$ . Deriving the expectation value of the new parity operator, $e^{i\pi \sum \hat{n}_i}$, where subscript $i$ denotes the different subsystems, we obtain 
\begin{align*}
    \langle e^{i\pi \sum \hat{n}_i} \rangle &= Tr[e^{i\pi \hat{n}_1}\rho_{coh} \otimes e^{i\pi \hat{n}_2}\rho_{env}] \\ \nonumber
    &=Tr[\bra{\alpha}e^{i\pi \hat{n}_1}\ket{\alpha} \otimes e^{i\pi \hat{n}_2}\rho_{env} ] \\ \nonumber
    &= Tr[e^{-2\Bar{n}} \otimes e^{i\pi \hat{n}_2}\rho_{env} ] \\ \nonumber
    &= e^{-2\Bar{n}}\langle \hat{\Pi} \rangle_{env},
\end{align*}

where $\langle \hat{\Pi} \rangle_{env}$ is bounded between 1 and -1. For large enough $\Bar{n}$, the whole expectation value goes to zero regardless of the form of $\rho_{env}$.

\subsubsection{Loss and detector inefficiency}
Consider an imperfect detector with quantum efficiency $\eta<1$. This can be modeled by placing a fictitious `loss beamsplitter' with reflectivity $r=\sqrt{1-\eta}$ and transmittivity $t=\sqrt{\eta}$ such that $r^2+t^2=1$ in front of a perfect detector and performing a partial trace over the reflected mode. The beamsplitter operator acting on bosonic modes $a$ and $b$ is given by 
\begin{equation}
    \hat{B}_{ab}=e^{\theta(\hat{a}\hat{b}^\dag-\hat{a}^\dag\hat{b})},
\end{equation}
where $r=\cos\theta$, $t=\sin\theta$. Sending a coherent state, $\ket{\alpha}$, to an imperfect detector is then the same as sending the density operator
\begin{align}
\label{eq:coherent_loss}
    \rho&=\text{Tr}_b\left[\hat{B}_{ab}\left(\ket{\alpha}\bra{\alpha}\right)_a \otimes\left(\ket{0}\bra{0}\right)_b\hat{B}^\dag_{ab} \right] \\
    &=\text{Tr}_b\left[\left(\ket{\sqrt{\eta}\alpha}\bra{\sqrt{\eta}\alpha}\right)_a \otimes\left(\ket{\sqrt{1-\eta}\alpha}\bra{\sqrt{1-\eta}\alpha}\right)_b \right]\\
    &=\left(\ket{\sqrt{\eta}\alpha}\bra{\sqrt{\eta}\alpha}\right)_a
\end{align}
to a perfect detector. Thus, for coherent states, all measurements made with PNRDs having $\eta<1$ can instead be treated as ideal detectors where the measured state is just a different coherent state.

\subsubsection{Unbalanced splitting and efficiency}
Suppose we send the coherent state $\ket{\alpha}$ to our three-detector system. Due to unbalanced splitting between different paths or small variations in detector efficiency, each TES may see a different signal. Together, the statistics of the photon number summed across all three channels will still be that of a coherent state but with potentially different effective amplitude. 

For an input coherent state and vacuum in the unused beamsplitter ports, $\ket{\alpha}_a\ket{0}_b\ket{0}_c$, the beamsplitter system shown in Fig.~\ref{fig:exp}(a) transforms the state to
\begin{equation}
    \hat{B}_{ac}\hat{B}_{ab}\ket{\alpha}_a\ket{0}_b\ket{0}_c=\ket{t_1t_2\alpha}_a\ket{r_1\alpha}_b\ket{t_1r_2\alpha}_c,
\end{equation}
where $r_k,t_k$ are the beamsplitter coefficients for beamsplitter $k$. Suppose now that the three detectors have quantum efficiencies $\eta_a,\eta_b$, and $\eta_c$. Using Eq.~\ref{eq:coherent_loss} for each mode, the effective state sent to three perfect detectors is then
\begin{align}
    \ket{\psi}&=\ket{\beta_a}_a\ket{\beta_b}_b\ket{\beta_c}_c\\
    &=e^{-\tfrac12\lvert\beta_a\beta_b\beta_c\rvert^2}\sum^\infty_{n_a=0}\sum^\infty_{n_b=0}\sum^\infty_{n_c=0}\frac{\beta_a^{n_a}\beta_b^{n_b}\beta_c^{n_c}}{\sqrt{n_a!n_b!n_c!}}\ket{n_a}_a\ket{n_b}_b\ket{n_c}_c
\end{align}
where
\begin{align}
    \beta_a&=\sqrt{\eta_a}t_1t_2\alpha, \\
    \beta_b&=\sqrt{\eta_b}r_1\alpha, \\
    \beta_c&=\sqrt{\eta_c}t_1r_2\alpha.
\end{align}
The probability to measure the total photon number summed across all detectors, $m=n_a+n_b+n_c$, is given by 
\begin{align}
P(m)&=e^{-\lvert\beta_a\beta_b\beta_c\rvert^2}\sum^m_{n_a=0}\sum^{m-n_a}_{n_b=0}\frac{\lvert\beta_a\rvert^{2n_a}\lvert\beta_b\rvert^{2n_b}\lvert\beta_c\rvert^{2(m-n_a-n_b)}}{n_a!n_b!(m-n_a-n_b)!}\\
&=e^{-\lvert\beta_a\beta_b\beta_c\rvert^2}\frac{\left(\lvert\beta_a\rvert^2+\lvert\beta_b\rvert^2+\lvert\beta_c\rvert^2\right)^m}{m!},
\end{align}
which is the same probability distribution that would be obtained by measuring a coherent state of amplitude $\alpha'=\sqrt{\lvert\beta_a\rvert^2+\lvert\beta_b\rvert^2+\lvert\beta_c\rvert^2}$ with a single detector of efficiency $\eta=1$.
\subsection{Experimental Methods}

The coherent state sent to the PNRD is generated by pulsing a continuous-wave 1064 nm laser using an acousto-optical modulator (AOM) as an optical switch. The pulse duration is set to be less than 100 ns, which is well within the rising-edge time of the detection signal. The pulses are sent at a repetition rate of 12.5 kHz to ensure that the detector has re-cooled and thermal noise is at a minimum. This rate can be increased to 50 kHz without incurring substantial ill-effects. Each split pulse is coupled to a TES channel through standard single-mode optical fiber. Details on TES operation within a cryostat can be found in Refs.~\cite{Gerrits2016,Nehra2019}.  In this work, we additionally filter the output signal to remove the DC component and implement a low-noise external amplifier to bring the signal to within a 500 mV range.

\subsubsection{Data acquisition}

The amplified output signal is sent to a custom-built Ethernet-based
flash analog-to-digital converter (EFADC) capable of collecting and processing TES signals for up to 8 channels. The device is based on a field-programmable gate array (FPGA) which samples a signal with 12-bit resolution at a rate of 250 MHz. The internal memory and processing speed allow the device to collect up to 32 µs worth of signal points, perform rudimentary calculations on the data to determine key parameters, and transfer the calculated parameters to a hard disk all before the next signal pulse arrives. 

The EFADC is triggered by an external pulse signal corresponding to the arrival time of each coherent state pulse. If the incoming signal rises above a user-defined noise threshold, the EFADC begins integrating waveform until the signal falls below a second threshold that can be set to account for hysteresis. The integrated signal area, maximum peak height, signal duration, timestamp of signal start, and timestamp of signal maximum are all recorded. All parameters can be used for additional signal characterization in post-processing, but we find that pulse area is sufficient to achieve large photon-number resolution.

\subsubsection{Efficiency calibration}
Transition-edge sensors have managed to reach up to $98\%$ quantum efficiency (QE)~\cite{Fukuda2011}, but it is important to characterize the precise response of our detection system at 1064 nm. The power in a given pulse sent to each TES detector is on the order of several pW, so care must be taken to accurately calibrate the QE. First, we constructed and characterized a high-amplification photodetection circuit with a low-power sensitivity threshold at approximately 200 pW. Calibration for this detector was based on a Scientech pyroelectric calorimeter and a series of precision attenuators. The home-build photodetector was then used in conjunction with the attenuators to calibrate each TES channel individually. Laser-light was split at a 95:5 beamsplitter where the stronger portion was sent to the photodetector and the weaker portion was further attenuated and sent to the TES. This calibrated attenuation included the effects of imperfect fiber coupling so the TES quantum efficiency could be directly measured.

For each detector, $10^6$ pulses were sent simultaneously to the photodetector and the TES channel under test. The mean photon number was extracted from the PNRD and compared with the classical signal power to determine the QE. We measured a QE of $97(5)\%$ for Channel 1, $93(5)\%$ for Channel 2, and $91(5)\%$ for Channel 3. The $5\%$ uncertainly originates from the absolute error on the Scientech pyroelectric calorimeter, uncertainty on splitting ratio, and error on the attenuation calibration. All channels used were thus measured to have a QE above $90\%$.
\subsubsection{Phase randomization}
ED~1 in the Extended Data shows the randomness tests for data where phase noise has been introduced to the coherent state. This is achieved by driving a mirror-mounted piezoelectric actuator (PZT) to change the optical path length over a range of one wavelength, or 1064 nm. The PZT was driven with a 100 Hz triangle-wave function, which was chosen to be much slower than the pulse repetition rate to ensure all phases over the range from $0$ to $2\pi$ were equally represented amongst the entire data set.

\subsubsection{Randomness characterization}

Here we will follow the work detailed in \cite{Gerry2022} on how the photon-number counts were binned to generate multiplicatively longer bit sequences as well as how the bit sequence was tested for randomness.  We start with the case of mod$(2)$ binning, in which each detection event corresponds to an outcome of even(0) or odd(1), the measurement probabilities are given by 
\begin{equation}
    P^{(2)}_{0(1)} = \langle \hat{P}^{(2)}_{0(1)}\rangle = \tfrac{1}{2}\left(1\pm e^{-2\bar{n}} \right)\;\;\to\;\;P^{(2)}_{k}=\tfrac{1}{2}\left(1+\left(-1\right)^k e^{-2\bar{n}}\right),
\end{equation}
where $\bar{n}$ is the average photon number of the coherent state and 
\begin{equation}
    \hat{P}^{(2)}_{k} = \sum_{m=0}^{\infty}\ket{2m+k}\bra{2m+k},
\end{equation}
are the even (k=0) and odd (k=1) projection operators. For large average photon numbers, the balancement between even/odd probabilities is maintained (i.e. $e^{-2\bar{n}}\to 0$). In terms of these projectors, the corresponding parity operator is given by $\hat{\Pi}=\hat{P}^{(2)}_{0}-\hat{P}_{1}^{(2)}$. Similarly, we can define projectors for the case of mod(4) binning  

\begin{equation}
    \hat{P}^{(4)}_{k} = \sum_{m=0}^{\infty}\ket{4m+k}\bra{4m+k},
\end{equation}
where each mod(2) bin is further broken down into bins containing every other even/odd photon count.  For example, the $k=0$ bin is comprised of the photon number counts $\{0,4,8,... \}$ while the $k=2$ bins counts $\{2,6,10,..\}$ and likewise for the odd counts. In this sense, mod(4) binning is akin to a higher-order parity measurement. It is clear then that the parity operator can be expressed as 
\begin{equation}
    \hat{\Pi} = \hat{P}^{(4)}_{0}+\hat{P}^{(4)}_{2}- \left(\hat{P}^{(4)}_{1}+\hat{P}^{(4)}_{3}\right) \equiv \hat{P}^{(2)}_{0}-\hat{P}^{(2)}_{1},
\end{equation}
and the binning probabilities are in turn given by
\begin{align}
    P_{k}^{(4)} = \langle \hat{P}_{k}^{(4)} \rangle &= e^{-\bar{n}}\sum_{n=0}^{\infty}\frac{\bar{n}^{4n+k}}{\left(4n+k\right)!} \nonumber \\
	&= \frac{1}{4}\left(1+2e^{-\bar{n}}\cos\left(\bar{n}-\tfrac{k\pi}{2}\right)+\left(-1\right)^{k}e^{-2\bar{n}}\right).
	\label{eqn:mod4_prob}
\end{align}
The length of the bit sequence can then be made longer by taking the remainders and mapping them to the dual-bit values according to $\left\{0,1,2,3\right\}\to\left\{00,01,10,11\right\}$.  This same form of mapping holds for higher modulo binning.  Note the largest biasing term in Eq.~\ref{eqn:mod4_prob} is larger than the mod(2) biasing term by a square root.  This implies a trade-off when binning the data: larger bit sequence generation comes at the cost of requiring a higher coherent state average photon number.  This procedure can be generalized for mod(Q) where the projectors are given by 
\begin{equation}
    \hat{P}^{(Q)}_{k}=\sum_{m=0}^{\infty}\ket{Qm+k}\bra{Qm+k},
\end{equation}
and the corresponding parity operator can in turn be constructed as 
\begin{equation}
    \hat{\Pi} = \sum_{k=0}^{Q-1}(-1)^k\hat{P}_{k}^{(Q)} \equiv \hat{P}_{0}^{(2)} - \hat{P}_{1}^{(2)}.
\end{equation}
The tested data is based off of 107911769 photon-number counts from a coherent source of average photon number $\bar{n}\approx57$.  For a trial size of $7.5\times10^{5}$, this corresponds to $n=\{143,287,431,575,719\}$ trials for mod$\{2,4,8,16,32\}$, respectively. We subject this data to a suite of randomness tests outlined by NIST SP800-22 \cite{ref:NIST_randomness_tests} in order to demonstrate that the generated bit sequence is truly random. We note that our methodology for determining randomness is the same employed in testing the randomness of bit sequences generated using the protocols of the NIST encryption standard competition finalists, detailed in Soto \textit{et al.} \cite{ref:nistSoto}, utilized in the verification of new randomness tests by Do{\u g}anaksoy \textit{et al.} \cite{ref:Doganaksoy} and implemented in the cryptographically-secure Intrinsic ID Zign software-based RNG \cite{schrijencreating}.  In ED~2 we plot the results of these tests for mod$\{2,4,16,32\}$. Note the mod(8) result can be found within the main body text. Due to the large number of tests available for judging whether a sequence is random or not, there is no `complete' or systematic approach to proving randomness. Instead, one relies on providing sufficient evidence that a given sequence is indeed random. For each trial, a series of tests are performed and a $P$-value is obtained for each test  corresponding to the probability that a perfect random number generator would produce a sequence \textit{less} random than the sequence being tested.  If this $P$-value is greater than the chosen significance level of $\alpha=0.01$ (1\%), the test is considered passed (successful) and the trial is accepted as random.  The proportion is then defined as the ratio of successful trials to the total number of trials (i.e. the success rate). Included in our analysis is the confidence interval (CI), i.e. the range of estimation for the success rate of a particular test given a 99\% confidence level. Typically, the CI for a set of Bernoulli trials with a success rate of $\hat{p}$ can be fairly approximated by that of the normal distribution
\begin{equation}
\text{CI} \approx \hat{p} \pm z\sqrt{\frac{\hat{p}\left(1-\hat{p}\right)}{n}}, 
\end{equation}
where $n$ is the total number of trials and $z$ is the $1-\tfrac{\alpha}{2}$ quantile probit function (i.e. the inverse cumulative distribution function for the normal distribution). However, this approximation to the binomial distribution, which is more representative of a set of Bernoulli trials, is only valid when the number of trials is on the order of $n\gtrsim 10^{4}$ and/or where the success rates are sufficiently far away from the boundary values of $0,1$. This proves to be an insufficient approximation for our data.  We instead turn to the asymmetric Wilson score approximation \cite{ref:Wilson} to the normal distribution given by
\begin{equation}
\text{CI}_{\text{ws}} = \frac{n}{n+z^2}\left(\hat{p}+\frac{z^2}{2n}\right)\pm\frac{zn}{n+z^2}\sqrt{\frac{\hat{p}\left(1-\hat{p}\right)}{n}+\frac{z^2}{4n^2}}.
\end{equation}
The Wilson score confidence interval, CI$_{\text{ws}}$, for a 99\% confidence level are represented by horizontal dashed blue lines in Fig.~\ref{fig:mod8randomness}, ED~1 and ED~2. In addition, we plot for each test the equivalent definition of the CI$_{\text{ws}}$
\begin{equation}
\text{CI}_{\text{ws}} = \frac{n_s+\tfrac{1}{2}z^2}{n+z^2} \pm \frac{z}{n+z^2}\sqrt{\frac{n_sn_f}{n}+\frac{z^2}{4}},
\label{eqn:WI_interval}
\end{equation}
where $n_s,n_f=n-n_s$ are the number of successful and failed trials, respectively. The success rate is then given by $\hat{p}=n_s/n$.  This measure provides a range for each test in which the mean proportion is likely to fall given repeated testing of the bit generation method (i.e. more trials performed) and is represented by red error bars in Fig.~\ref{fig:mod8randomness}, ED~1 and ED~2. Sufficient evidence of randomness exists if the proportion lies above the lower bound of the $\text{CI}_{\text{ws}}$ for all tests considered.  By this criterion, we conclude that the generated bit sequences for the cases of mod$\left\{2,4,8\right\}$ binning are random while the generated bit sequences for mod$\left\{16,32,..\right\}$ binning are not random.

To further validate our results, we reiterate that for the case of a coherent state with average photon number $\bar{n}\approx57$, we expect the balancement of binning probabilities to hold for up to mod(8) binning. Higher modulo binning will introduce larger degrees of bias into the binning probabilities, as seen in Eq.~\ref{eqn:mod4_prob}. An approximate trend is that the largest biasing term in the binning probabilities for the case of mod($Q$) binning is $\;\propto \text{exp}\left(-\tfrac{4\bar{n}}{Q}\right)$, such that if one wanted to maintain the same degree of bias as the mod(2) binning case, one would need a coherent state with an average photon number $\tfrac{1}{2}Q$ times larger. For a static $\bar{n}$, higher mod binning will subsequently result in a generated bit sequence that does not display randomness as there will be a significant amount of bias in the higher-modulo binning probabilities.  For reference, the impact of bias on the randomness of the bit sequence is reflected in ED~2, where as predicted the mod(16) and mod(32) binning cases show evidence that the generated bit sequence is \textit{not} random since for both cases several test proportions fall outside of the CI$_{\text{ws}}$.  Even more specifically, only a few tests fail for the mod(16) case and most fail for the mod(32), reflecting that more bias is introduced as a function of the modulo binning size. Likewise, this also further strengthens the argument that the mod$\left\{2,4,8\right\}$ cases result in a random bit sequence, as our experimental data align perfectly with theoretical predictions.

\subsection{Additional Data}
Further analyses of experimental data are shown in the Extended Data. Full characterization of the randomness tests on all data is shown in ED~1 and~2. The effect of error-rate reduction through binning modifications is shown in ED~3 with the normalized Gaussian fitting for all three TES channels displayed in ED~4. Specific error rates for different photon-number measurements on each channel based on different histogram binning are shown in ED~5. Theoretical residual bias for photon-number measurements modulo $d$ with an upper limit of 100 resolvable photons are shown in ED~6.

\section{Data Availability}
The data supporting plots within this paper are available at \url{https://doi.org/10.6084/m9.figshare.21304524.v1} and \url{https://doi.org/10.6084/m9.figshare.21291318}.
Additional data used for detector calibration can be obtained from the corresponding authors on reasonable request.

\section{Code Availability}
The codes used to process and analyze the data can be obtained from the corresponding authors on reasonable request.

\pagebreak

\clearpage
\newpage
\widetext
\section{Extended Data} \label{sec:ExtraData}

\begin{figure*}[h]
    \centering
    \includegraphics[width = 0.55\textwidth]{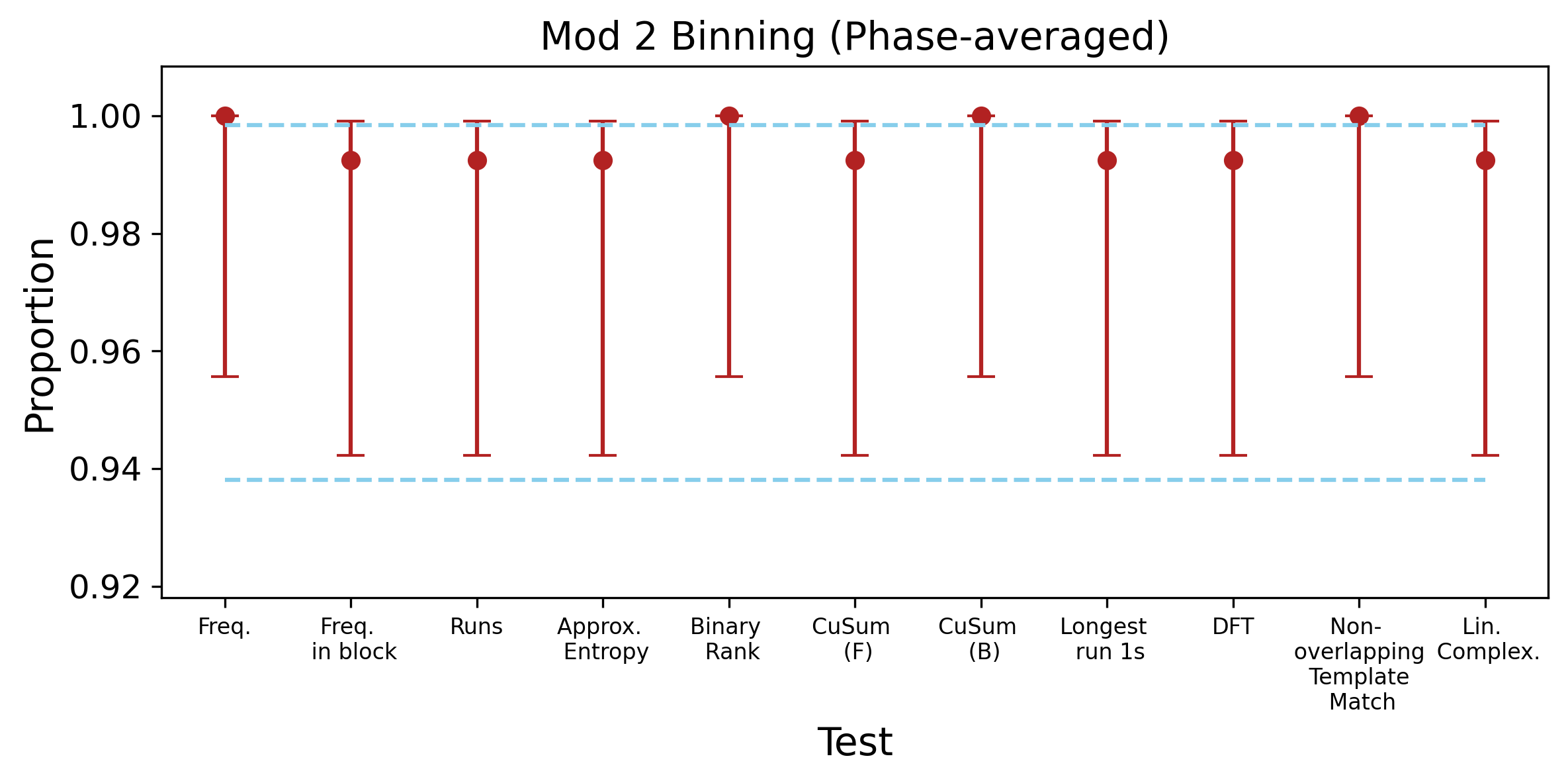}
\caption{Randomness tests for bit strings obtained from modulo 2 binning the sampled photon number from a mixture of coherent states with randomized phase. All tests pass indicating phase stability has no bearing on the quality of QRNG. The error bars for each proportion are computed from the Wilson score interval of Eq.~\ref{eqn:WI_interval} where $n=143$ is the total number of trials and $n_s\;\left(n_f\right)$ are the number of successful (failed) trials for a significance level of $\alpha=0.01$. Given repeated testing of the bit generation method, the error bars denote the probability range for which the proportion is likely to fall.
}
    \label{fig:phase_rand_tests}
\end{figure*}

\begin{figure*}
    \centering
    \includegraphics[width = 0.9\textwidth]{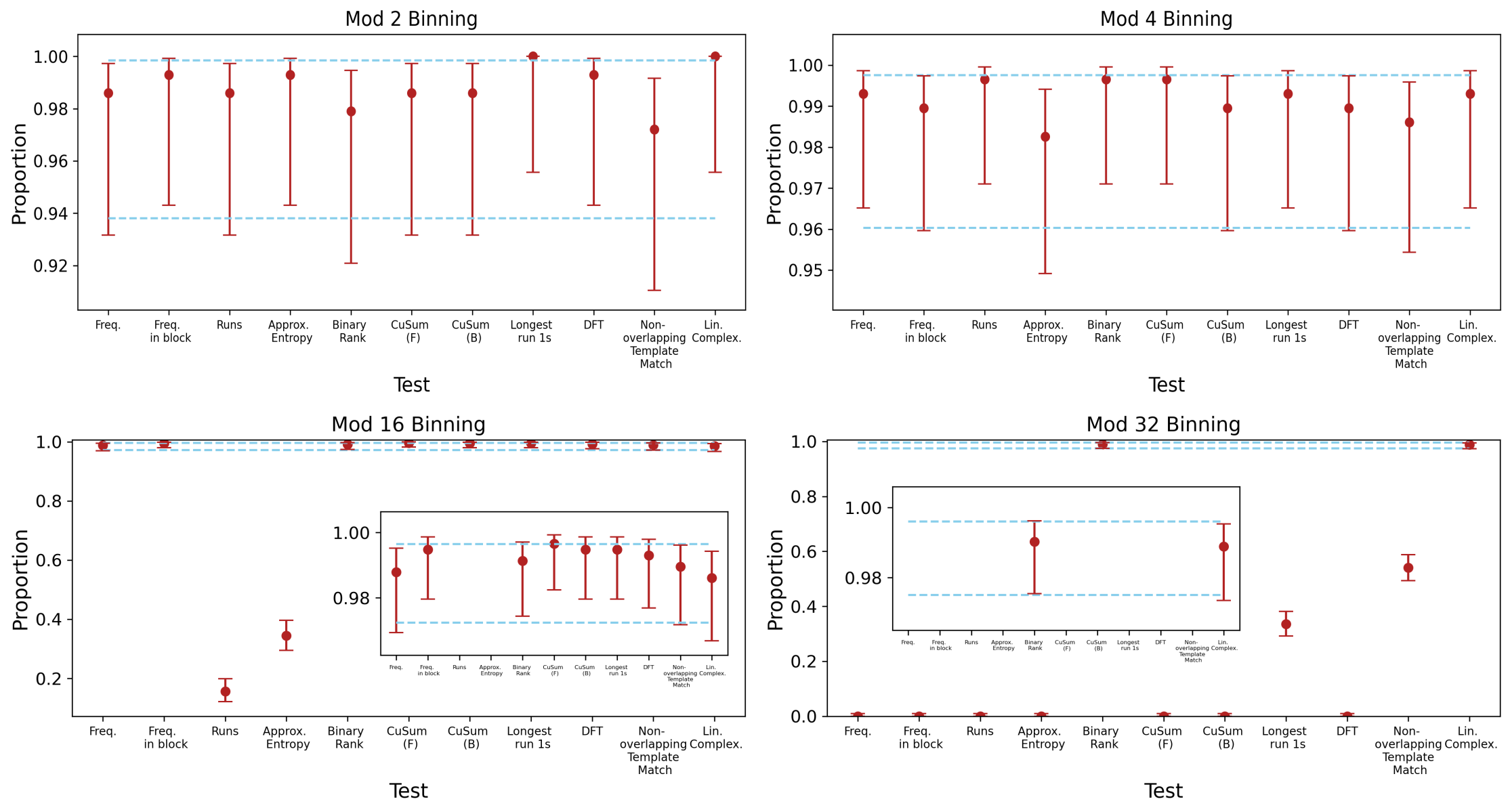}
\caption{Randomness tests for the resultant bit strings based on how the measured data is binned (Mod 8 data shown in the main text). Mod 2, Mod 4, and Mod 8 tests all indicate randomness, while some tests begin to fail for Mod 16 and Mod 32. This is expected due to the non-zero residual biases for a coherent state distribution with mean photon number $\Bar{n}=57$ and a PNRD limit of 100 photons. The error bars for each proportion are computed from the Wilson score (confidence) interval of Eq.~\ref{eqn:WI_interval} where $n=\left\{143,287,575,719\right\}$ is the total number of trials for mod$\left\{2,4,16,32\right\}$ binning, respectively, and $n_s\;\left(n_f\right)$ are the number of successful (failed) trials for a significance level of $\alpha=0.01$. Given repeated testing of the bit generation method, the error bars denote the probability range for which the proportion is likely to fall.
}
    \label{fig:randomness_tests}
\end{figure*}

\begin{figure*}
    \centering
    \includegraphics[width = 0.8\textwidth]{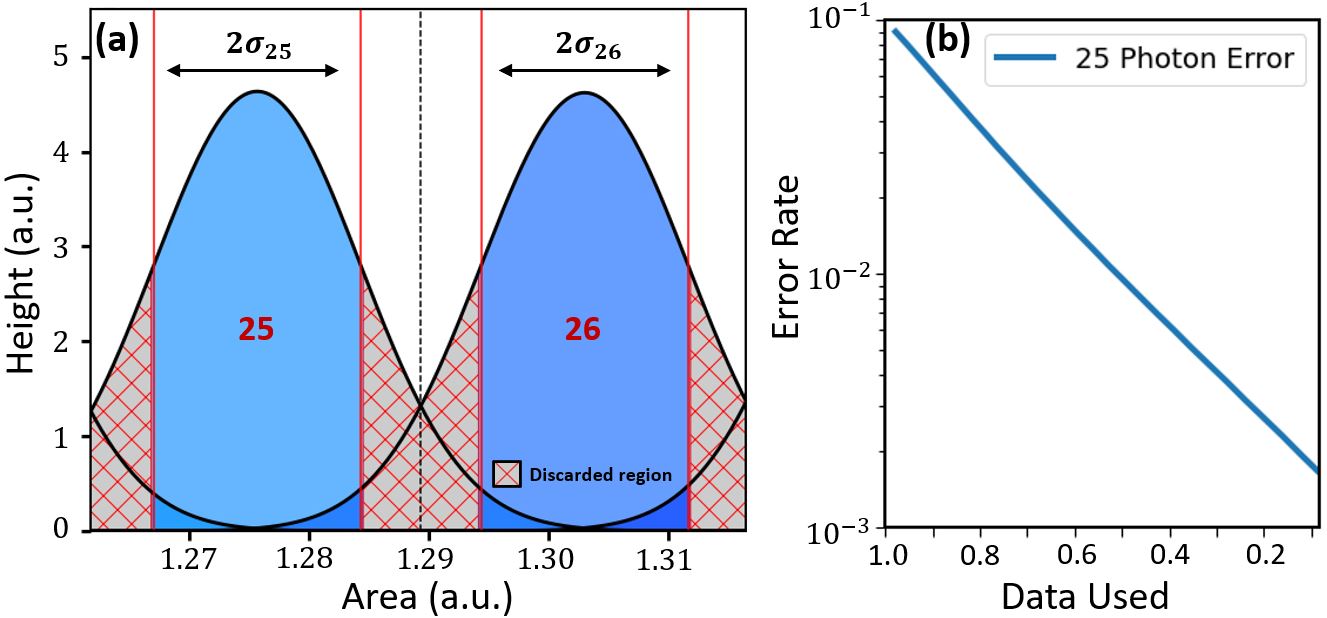}
\caption{Error-rate reduction on photon-number resolution through post-selection of data. (a) By excluding data points with measured areas further from the center of each bin, the portion of overlap from neighboring Gaussians can be substantially reduced. The location of the new binning thresholds must be the same fraction of the Gaussian peak width, $\sigma_n$, for each bin. Here, $2\sigma_n$ is chosen. (b) Error rate to incorrectly characterize a true 25 photon event as a function of the proportion of measurement data kept. 
}
    \label{fig:cut_binerror}
\end{figure*}

\begin{figure*}
    \centering
    \includegraphics[width = \textwidth]{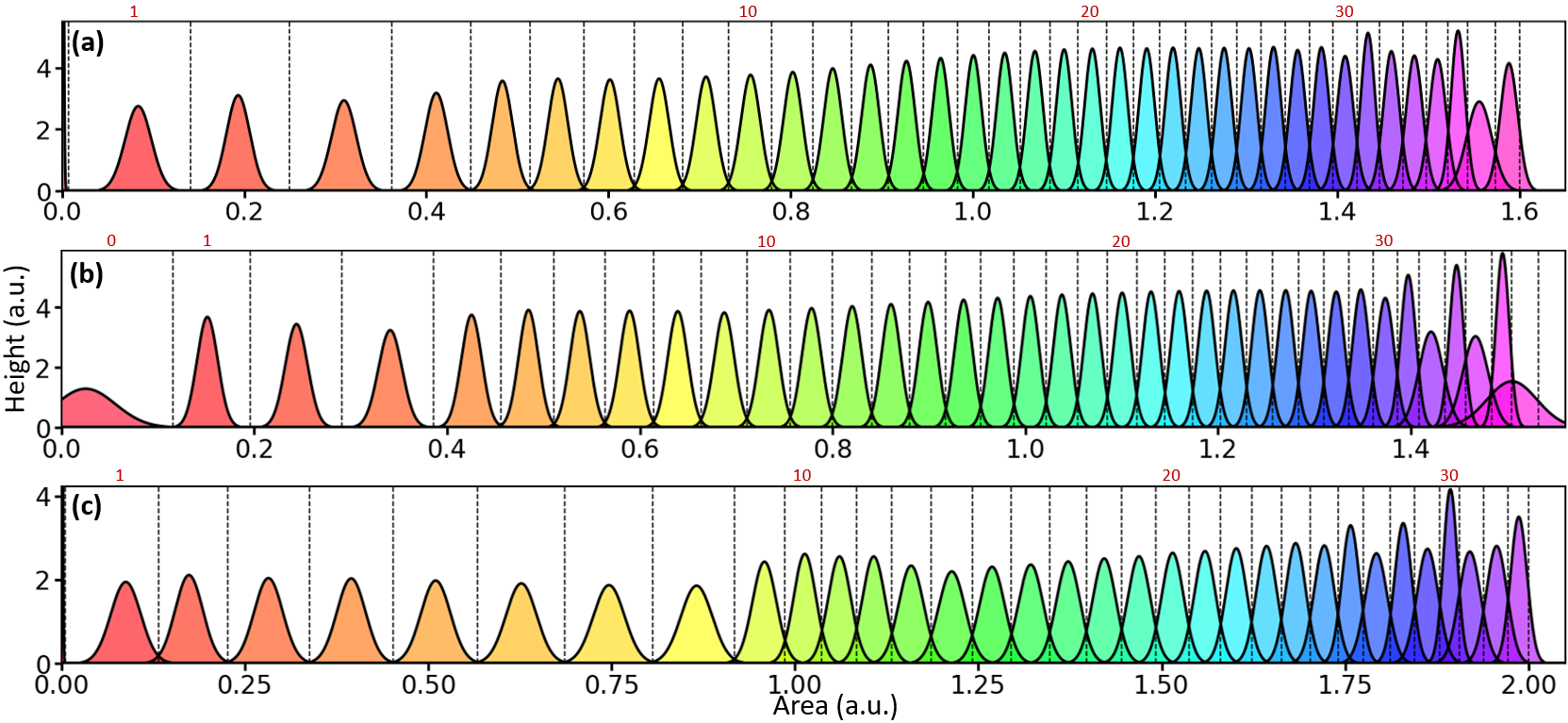}
\caption{Normalized Gaussian fits for the histogrammed area measurements  TES channel 1 (a), 2 (b), and 3 (c). Note that for channels 1 and 3, the FPGA thresholds are set above the electronics noise such that zero photon events have a measured area of zero. For channel 2, electronics noise can drift slightly above the set voltage threshold so that small, non-zero areas are recorded for zero photon events.
}
    \label{fig:gauss_ovlp}
\end{figure*}

\begin{figure*}
    \centering
    \includegraphics[width = \textwidth]{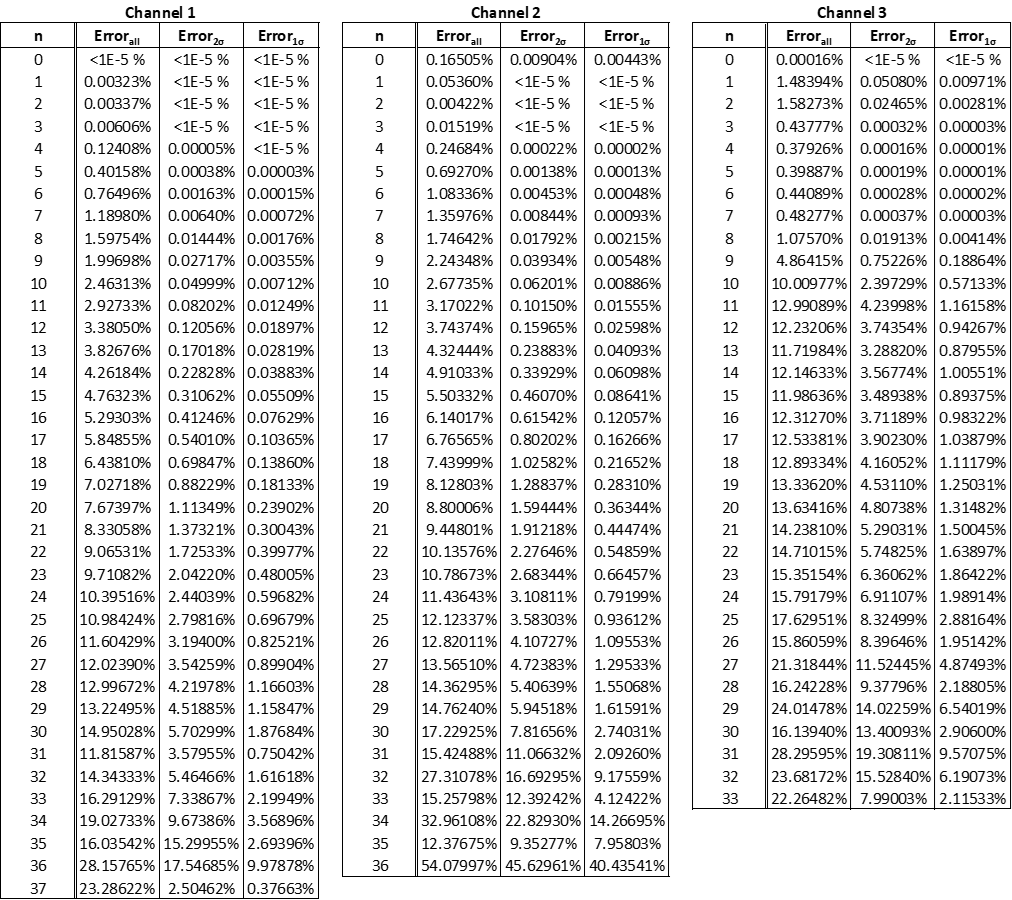}
\caption{Error rates for all detection channels depending on binning. Error percentages indicate the probability to incorrectly count a measurement that was a true $n$ photon event. $\text{Error}_{all}$ includes all areas and uses the Gaussian intersections to place bins. $\text{Error}_{2\sigma}$ discards area events occurring outsides of a $2\sigma$ width centered around each Gaussian in the histogram fit. The thrown-out events account for $32\%$ of all measurements. The $\text{Error}_{1\sigma}$ discards area events occurring outsides of a $1\sigma$ width centered around each Gaussian in the histogram fit. This removes $62\%$ of the measured data but drastically reduces counting errors.
}
    \label{fig:error_rates}
\end{figure*}

\begin{figure*}
    \centering
    \includegraphics[width = \textwidth]{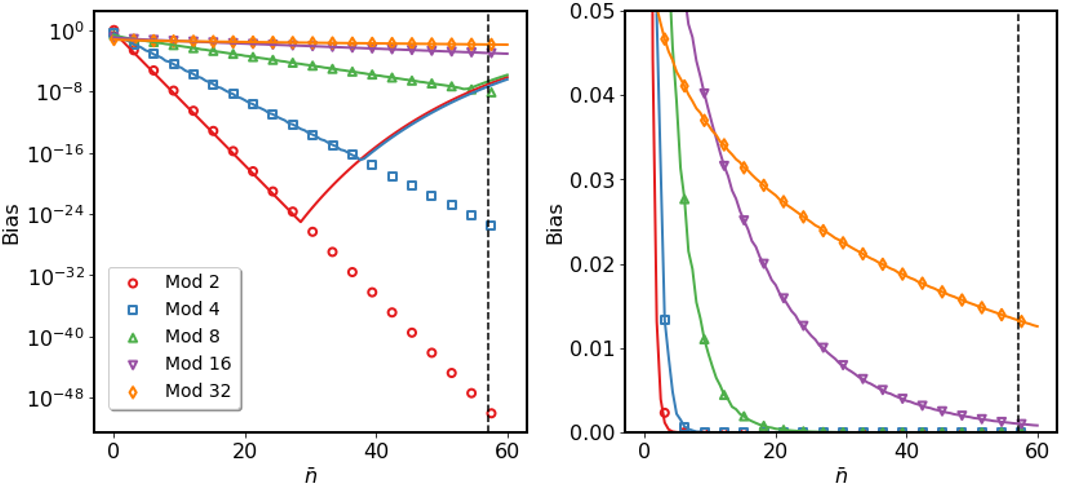}
\caption{Residual bias based on modulo binning of a photon number distribution for coherent state of mean photon number $\Bar{n}$. Markers indicate the theoretical deviation from a uniformly random distribution if one had infinite photon-number resolving capability while solid lines give the expected bias with a truncation of the photon number distribution beyond 100 photons. The vertical dashed line indicates a coherent state with $\Bar{n}=57$ such as used in this experiment where the residual bias for mod 2, mod 4, and mod 8 binning are the same. The two plots are identical with the plot at left showing log scale.
}
    \label{fig:residual_bias}
\end{figure*}



\begin{thebibliography}{10}
\expandafter\ifx\csname url\endcsname\relax
  \def\url#1{\texttt{#1}}\fi
\expandafter\ifx\csname urlprefix\endcsname\relax\def\urlprefix{URL }\fi
\providecommand{\bibinfo}[2]{#2}
\providecommand{\eprint}[2][]{\url{#2}}

\bibitem{Einstein1905_2}
\bibinfo{author}{Einstein, A.}
\newblock \bibinfo{title}{On a heuristic point of view about the creation and
  conversion of light}.
\newblock \emph{\bibinfo{journal}{Annalen der Physik}}
  \textbf{\bibinfo{volume}{17}}, \bibinfo{pages}{132--148}
  (\bibinfo{year}{1905}).

\bibitem{Salart2008}
\bibinfo{author}{Salart, D.}, \bibinfo{author}{Baas, A.},
  \bibinfo{author}{Branciard, C.}, \bibinfo{author}{Gisin, N.} \&
  \bibinfo{author}{Zbinden, H.}
\newblock \bibinfo{title}{Testing the speed of ‘spooky action at a
  distance’}.
\newblock \emph{\bibinfo{journal}{Nature}} \textbf{\bibinfo{volume}{454}},
  \bibinfo{pages}{861--864} (\bibinfo{year}{2008}).

\bibitem{Gisin2007}
\bibinfo{author}{Gisin, N.} \& \bibinfo{author}{Thew, R.}
\newblock \bibinfo{title}{Quantum communication}.
\newblock \emph{\bibinfo{journal}{Nature Photonics}}
  \textbf{\bibinfo{volume}{1}}, \bibinfo{pages}{165} (\bibinfo{year}{2007}).
\newblock \urlprefix\url{https://doi.org/10.1038/nphoton.2007.22}.

\bibitem{becerra2013experimental}
\bibinfo{author}{Becerra, F.} \emph{et~al.}
\newblock \bibinfo{title}{Experimental demonstration of a receiver beating the
  standard quantum limit for multiple nonorthogonal state discrimination}.
\newblock \emph{\bibinfo{journal}{Nature Photonics}}
  \textbf{\bibinfo{volume}{7}}, \bibinfo{pages}{147--152}
  (\bibinfo{year}{2013}).

\bibitem{Slussarenko2017}
\bibinfo{author}{Slussarenko, S.} \emph{et~al.}
\newblock \bibinfo{title}{Unconditional violation of the shot-noise limit in
  photonic quantum metrology}.
\newblock \emph{\bibinfo{journal}{Nature Photonics}}
  \textbf{\bibinfo{volume}{11}}, \bibinfo{pages}{700--703}
  (\bibinfo{year}{2017}).

\bibitem{Nehra2019}
\bibinfo{author}{Nehra, R.} \emph{et~al.}
\newblock \bibinfo{title}{State-independent quantum state tomography by
  photon-number-resolving measurements}.
\newblock \emph{\bibinfo{journal}{Optica}} \textbf{\bibinfo{volume}{6}},
  \bibinfo{pages}{1356--1360} (\bibinfo{year}{2019}).
\newblock
  \urlprefix\url{http://www.osapublishing.org/optica/abstract.cfm?URI=optica-6-10-1356}.

\bibitem{Zhong2020}
\bibinfo{author}{Zhong, H.-S.} \emph{et~al.}
\newblock \bibinfo{title}{Quantum computational advantage using photons}.
\newblock \emph{\bibinfo{journal}{Science}} \textbf{\bibinfo{volume}{370}},
  \bibinfo{pages}{1460--1463} (\bibinfo{year}{2020}).

\bibitem{Arrazola2021}
\bibinfo{author}{Arrazola, J.} \emph{et~al.}
\newblock \bibinfo{title}{Quantum circuits with many photons on a programmable
  nanophotonic chip}.
\newblock \emph{\bibinfo{journal}{Nature}} \textbf{\bibinfo{volume}{591}},
  \bibinfo{pages}{54--60} (\bibinfo{year}{2021}).

\bibitem{campbell2016recent}
\bibinfo{author}{Campbell, J.~C.}
\newblock \bibinfo{title}{Recent advances in avalanche photodiodes}.
\newblock \emph{\bibinfo{journal}{Journal of Lightwave Technology}}
  \textbf{\bibinfo{volume}{34}}, \bibinfo{pages}{278--285}
  (\bibinfo{year}{2016}).

\bibitem{Becerra2015}
\bibinfo{author}{Becerra, F.}, \bibinfo{author}{Fan, J.} \&
  \bibinfo{author}{Migdall, A.}
\newblock \bibinfo{title}{Photon number resolution enables quantum receiver for
  realistic coherent optical communications}.
\newblock \emph{\bibinfo{journal}{Nature Photonics}}
  \textbf{\bibinfo{volume}{9}}, \bibinfo{pages}{48--53} (\bibinfo{year}{2015}).

\bibitem{Arrazola2019}
\bibinfo{author}{Arrazola, J.~M.} \emph{et~al.}
\newblock \bibinfo{title}{Machine learning method for state preparation and
  gate synthesis on photonic quantum computers}.
\newblock \emph{\bibinfo{journal}{Quantum Science and Technology}}
  \textbf{\bibinfo{volume}{4}}, \bibinfo{pages}{024004} (\bibinfo{year}{2019}).

\bibitem{Thekkadath2020}
\bibinfo{author}{Thekkadath, G.} \emph{et~al.}
\newblock \bibinfo{title}{Quantum-enhanced interferometry with large heralded
  photon-number states}.
\newblock \emph{\bibinfo{journal}{NPJ quantum information}}
  \textbf{\bibinfo{volume}{6}}, \bibinfo{pages}{1--6} (\bibinfo{year}{2020}).

\bibitem{Eaton2019}
\bibinfo{author}{Eaton, M.}, \bibinfo{author}{Nehra, R.} \&
  \bibinfo{author}{Pfister, O.}
\newblock \bibinfo{title}{Non-gaussian and gottesman--kitaev--preskill state
  preparation by photon catalysis}.
\newblock \emph{\bibinfo{journal}{New Journal of Physics}}
  \textbf{\bibinfo{volume}{21}}, \bibinfo{pages}{113034}
  (\bibinfo{year}{2019}).

\bibitem{Ra2020}
\bibinfo{author}{Ra, Y.-S.} \emph{et~al.}
\newblock \bibinfo{title}{Non-gaussian quantum states of a multimode light
  field}.
\newblock \emph{\bibinfo{journal}{Nature Physics}}
  \textbf{\bibinfo{volume}{16}}, \bibinfo{pages}{144--147}
  (\bibinfo{year}{2020}).

\bibitem{Walschaers2021}
\bibinfo{author}{Walschaers, M.}
\newblock \bibinfo{title}{Non-gaussian quantum states and where to find them}.
\newblock \emph{\bibinfo{journal}{PRX Quantum}} \textbf{\bibinfo{volume}{2}},
  \bibinfo{pages}{030204} (\bibinfo{year}{2021}).

\bibitem{Mari2012}
\bibinfo{author}{Mari, A.} \& \bibinfo{author}{Eisert, J.}
\newblock \bibinfo{title}{Positive wigner functions render classical simulation
  of quantum computation efficient}.
\newblock \emph{\bibinfo{journal}{Physical review letters}}
  \textbf{\bibinfo{volume}{109}}, \bibinfo{pages}{230503}
  (\bibinfo{year}{2012}).

\bibitem{bulmer2021boundary}
\bibinfo{author}{Bulmer, J.~F.} \emph{et~al.}
\newblock \bibinfo{title}{The boundary for quantum advantage in gaussian boson
  sampling}.
\newblock \emph{\bibinfo{journal}{Science Advances}}
  \textbf{\bibinfo{volume}{8}}, \bibinfo{pages}{eabl9236}
  (\bibinfo{year}{2021}).

\bibitem{furst2010high}
\bibinfo{author}{F{\"u}rst, H.} \emph{et~al.}
\newblock \bibinfo{title}{High speed optical quantum random number generation}.
\newblock \emph{\bibinfo{journal}{Optics express}}
  \textbf{\bibinfo{volume}{18}}, \bibinfo{pages}{13029--13037}
  (\bibinfo{year}{2010}).

\bibitem{ren2011quantum}
\bibinfo{author}{Ren, M.} \emph{et~al.}
\newblock \bibinfo{title}{Quantum random-number generator based on a
  photon-number-resolving detector}.
\newblock \emph{\bibinfo{journal}{Physical Review A}}
  \textbf{\bibinfo{volume}{83}}, \bibinfo{pages}{023820}
  (\bibinfo{year}{2011}).

\bibitem{Gerry2022}
\bibinfo{author}{Gerry, C.~C.} \emph{et~al.}
\newblock \bibinfo{title}{Proposal for a quantum random number generator using
  coherent light and a non-classical observable}.
\newblock \emph{\bibinfo{journal}{J. Opt. Soc. Am. B}}
  \textbf{\bibinfo{volume}{39}}, \bibinfo{pages}{1068--1074}
  (\bibinfo{year}{2022}).
\newblock
  \urlprefix\url{http://opg.optica.org/josab/abstract.cfm?URI=josab-39-4-1068}.

\bibitem{Lita2008}
\bibinfo{author}{Lita, A.~E.}, \bibinfo{author}{Miller, A.~J.} \&
  \bibinfo{author}{Nam, S.~W.}
\newblock \bibinfo{title}{Counting near-infrared single-photons with 95\%
  efficiency}.
\newblock \emph{\bibinfo{journal}{Opt. Expr.}} \textbf{\bibinfo{volume}{16}},
  \bibinfo{pages}{3032--3040} (\bibinfo{year}{2008}).

\bibitem{Fukuda2011}
\bibinfo{author}{Fukuda, D.} \emph{et~al.}
\newblock \bibinfo{title}{Titanium-based transition-edge photon number
  resolving detector with 98\% detection efficiency with index-matched
  small-gap fiber coupling}.
\newblock \emph{\bibinfo{journal}{Optics express}}
  \textbf{\bibinfo{volume}{19}}, \bibinfo{pages}{870--875}
  (\bibinfo{year}{2011}).

\bibitem{Gerrits2016}
\bibinfo{author}{Gerrits, T.}, \bibinfo{author}{Lita, A.},
  \bibinfo{author}{Calkins, B.} \& \bibinfo{author}{Nam, S.~W.}
\newblock \bibinfo{title}{Superconducting transition edge sensors for quantum
  optics}.
\newblock In \emph{\bibinfo{booktitle}{Superconducting devices in quantum
  optics}}, \bibinfo{pages}{31--60} (\bibinfo{publisher}{Springer},
  \bibinfo{year}{2016}).

\bibitem{Gerrits2012}
\bibinfo{author}{Gerrits, T.} \emph{et~al.}
\newblock \bibinfo{title}{Extending single-photon optimized superconducting
  transition edge sensors beyond the single-photon counting regime}.
\newblock \emph{\bibinfo{journal}{Optics Express}}
  \textbf{\bibinfo{volume}{20}}, \bibinfo{pages}{23798--23810}
  (\bibinfo{year}{2012}).

\bibitem{Harder2016}
\bibinfo{author}{Harder, G.} \emph{et~al.}
\newblock \bibinfo{title}{Single-mode parametric-down-conversion states with 50
  photons as a source for mesoscopic quantum optics}.
\newblock \emph{\bibinfo{journal}{Physical review letters}}
  \textbf{\bibinfo{volume}{116}}, \bibinfo{pages}{143601}
  (\bibinfo{year}{2016}).

\bibitem{Levine2012}
\bibinfo{author}{Levine, Z.~H.} \emph{et~al.}
\newblock \bibinfo{title}{Algorithm for finding clusters with a known
  distribution and its application to photon-number resolution using a
  superconducting transition-edge sensor}.
\newblock \emph{\bibinfo{journal}{J. Opt. Soc. Am. B}}
  \textbf{\bibinfo{volume}{29}}, \bibinfo{pages}{2066--2073}
  (\bibinfo{year}{2012}).
\newblock
  \urlprefix\url{http://josab.osa.org/abstract.cfm?URI=josab-29-8-2066}.

\bibitem{Morais2020}
\bibinfo{author}{Morais, L.~A.} \emph{et~al.}
\newblock \bibinfo{title}{Precisely determining photon-number in real-time}.
\newblock \emph{\bibinfo{journal}{arXiv:2012.10158 [physics.ins-det]}}
  (\bibinfo{year}{2020}).
\newblock \eprint{2012.10158}.

\bibitem{Gottesman2001}
\bibinfo{author}{Gottesman, D.}, \bibinfo{author}{Kitaev, A.} \&
  \bibinfo{author}{Preskill, J.}
\newblock \bibinfo{title}{Encoding a qubit in an oscillator}.
\newblock \emph{\bibinfo{journal}{Phys. Rev. A}} \textbf{\bibinfo{volume}{64}},
  \bibinfo{pages}{012310} (\bibinfo{year}{2001}).

\bibitem{Ghose2007}
\bibinfo{author}{Ghose, S.} \& \bibinfo{author}{Sanders, B.~C.}
\newblock \bibinfo{title}{Non-gaussian ancilla states for continuous variable
  quantum computation via gaussian maps}.
\newblock \emph{\bibinfo{journal}{J. Mod. Opt.}} \textbf{\bibinfo{volume}{54}},
  \bibinfo{pages}{855--869} (\bibinfo{year}{2007}).

\bibitem{ferrenberg1992monte}
\bibinfo{author}{Ferrenberg, A.~M.}, \bibinfo{author}{Landau, D.} \&
  \bibinfo{author}{Wong, Y.~J.}
\newblock \bibinfo{title}{Monte carlo simulations: Hidden errors from
  ‘‘good’’random number generators}.
\newblock \emph{\bibinfo{journal}{Physical Review Letters}}
  \textbf{\bibinfo{volume}{69}}, \bibinfo{pages}{3382} (\bibinfo{year}{1992}).

\bibitem{Ma2016}
\bibinfo{author}{Ma, X.}, \bibinfo{author}{Yuan, X.}, \bibinfo{author}{Cao,
  Z.}, \bibinfo{author}{Qi, B.} \& \bibinfo{author}{Zhang, Z.}
\newblock \bibinfo{title}{Quantum random number generation}.
\newblock \emph{\bibinfo{journal}{npj Quantum Information}}
  \textbf{\bibinfo{volume}{2}}, \bibinfo{pages}{1--9} (\bibinfo{year}{2016}).

\bibitem{herrero2017quantum}
\bibinfo{author}{Herrero-Collantes, M.} \& \bibinfo{author}{Garcia-Escartin,
  J.~C.}
\newblock \bibinfo{title}{Quantum random number generators}.
\newblock \emph{\bibinfo{journal}{Reviews of Modern Physics}}
  \textbf{\bibinfo{volume}{89}}, \bibinfo{pages}{015004}
  (\bibinfo{year}{2017}).

\bibitem{fujii2012thin}
\bibinfo{author}{Fujii, G.} \emph{et~al.}
\newblock \bibinfo{title}{Thin gold covered titanium transition edge sensor for
  optical measurement}.
\newblock \emph{\bibinfo{journal}{Journal of Low Temperature Physics}}
  \textbf{\bibinfo{volume}{167}}, \bibinfo{pages}{815--821}
  (\bibinfo{year}{2012}).

\bibitem{stefanov2000optical}
\bibinfo{author}{Stefanov, A.}, \bibinfo{author}{Gisin, N.},
  \bibinfo{author}{Guinnard, O.}, \bibinfo{author}{Guinnard, L.} \&
  \bibinfo{author}{Zbinden, H.}
\newblock \bibinfo{title}{Optical quantum random number generator}.
\newblock \emph{\bibinfo{journal}{Journal of Modern Optics}}
  \textbf{\bibinfo{volume}{47}}, \bibinfo{pages}{595--598}
  (\bibinfo{year}{2000}).

\bibitem{jennewein2000fast}
\bibinfo{author}{Jennewein, T.}, \bibinfo{author}{Achleitner, U.},
  \bibinfo{author}{Weihs, G.}, \bibinfo{author}{Weinfurter, H.} \&
  \bibinfo{author}{Zeilinger, A.}
\newblock \bibinfo{title}{A fast and compact quantum random number generator}.
\newblock \emph{\bibinfo{journal}{Review of Scientific Instruments}}
  \textbf{\bibinfo{volume}{71}}, \bibinfo{pages}{1675--1680}
  (\bibinfo{year}{2000}).

\bibitem{gabriel2010generator}
\bibinfo{author}{Gabriel, C.} \emph{et~al.}
\newblock \bibinfo{title}{A generator for unique quantum random numbers based
  on vacuum states}.
\newblock \emph{\bibinfo{journal}{Nature Photonics}}
  \textbf{\bibinfo{volume}{4}}, \bibinfo{pages}{711--715}
  (\bibinfo{year}{2010}).

\bibitem{sanguinetti2014quantum}
\bibinfo{author}{Sanguinetti, B.}, \bibinfo{author}{Martin, A.},
  \bibinfo{author}{Zbinden, H.} \& \bibinfo{author}{Gisin, N.}
\newblock \bibinfo{title}{Quantum random number generation on a mobile phone}.
\newblock \emph{\bibinfo{journal}{Physical Review X}}
  \textbf{\bibinfo{volume}{4}}, \bibinfo{pages}{031056} (\bibinfo{year}{2014}).

\bibitem{vonNeumann1951}
\bibinfo{author}{Von~Neumann, J.}
\newblock \bibinfo{title}{13. various techniques used in connection with random
  digits}.
\newblock \emph{\bibinfo{journal}{Appl. Math Ser}}
  \textbf{\bibinfo{volume}{12}}, \bibinfo{pages}{3} (\bibinfo{year}{1951}).

\bibitem{Peres1992rand}
\bibinfo{author}{Peres, Y.}
\newblock \bibinfo{title}{Iterating von neumann's procedure for extracting
  random bits}.
\newblock \emph{\bibinfo{journal}{The Annals of Statistics}}
  \bibinfo{pages}{590--597} (\bibinfo{year}{1992}).

\bibitem{zhao2008quantum}
\bibinfo{author}{Zhao, Y.}, \bibinfo{author}{Fung, C.-H.~F.},
  \bibinfo{author}{Qi, B.}, \bibinfo{author}{Chen, C.} \& \bibinfo{author}{Lo,
  H.-K.}
\newblock \bibinfo{title}{Quantum hacking: Experimental demonstration of
  time-shift attack against practical quantum-key-distribution systems}.
\newblock \emph{\bibinfo{journal}{Physical Review A}}
  \textbf{\bibinfo{volume}{78}}, \bibinfo{pages}{042333}
  (\bibinfo{year}{2008}).

\bibitem{cahall2017multi}
\bibinfo{author}{Cahall, C.} \emph{et~al.}
\newblock \bibinfo{title}{Multi-photon detection using a conventional
  superconducting nanowire single-photon detector}.
\newblock \emph{\bibinfo{journal}{Optica}} \textbf{\bibinfo{volume}{4}},
  \bibinfo{pages}{1534--1535} (\bibinfo{year}{2017}).

\bibitem{ref:Wilson}
\bibinfo{author}{Wilson, E.~B.}
\newblock \bibinfo{title}{Probable inference, the law of succession, and
  statistical inference}.
\newblock \emph{\bibinfo{journal}{Journal of the American Statistical
  Association}} \textbf{\bibinfo{volume}{22}}, \bibinfo{pages}{209--212}
  (\bibinfo{year}{1927}).
\newblock
  \urlprefix\url{https://www.tandfonline.com/doi/abs/10.1080/01621459.1927.10502953}.
\newblock
  \eprint{https://www.tandfonline.com/doi/pdf/10.1080/01621459.1927.10502953}.

\bibitem{Acin2016}
\bibinfo{author}{Ac{\'\i}n, A.} \& \bibinfo{author}{Masanes, L.}
\newblock \bibinfo{title}{Certified randomness in quantum physics}.
\newblock \emph{\bibinfo{journal}{Nature}} \textbf{\bibinfo{volume}{540}},
  \bibinfo{pages}{213--219} (\bibinfo{year}{2016}).

\bibitem{gerry2000heisenberg}
\bibinfo{author}{Gerry, C.~C.}
\newblock \bibinfo{title}{Heisenberg-limit interferometry with four-wave mixers
  operating in a nonlinear regime}.
\newblock \emph{\bibinfo{journal}{Physical Review A}}
  \textbf{\bibinfo{volume}{61}}, \bibinfo{pages}{043811}
  (\bibinfo{year}{2000}).

\bibitem{Marshall2015a}
\bibinfo{author}{Marshall, K.}, \bibinfo{author}{Pooser, R.},
  \bibinfo{author}{Siopsis, G.} \& \bibinfo{author}{Weedbrook, C.}
\newblock \bibinfo{title}{Quantum simulation of quantum field theory using
  continuous variables}.
\newblock \emph{\bibinfo{journal}{Phys. Rev. A}} \textbf{\bibinfo{volume}{92}},
  \bibinfo{pages}{063825} (\bibinfo{year}{2015}).
\newblock \urlprefix\url{https://link.aps.org/doi/10.1103/PhysRevA.92.063825}.

\end{thebibliography}

\begin{thebibliography}{10}
\expandafter\ifx\csname url\endcsname\relax
  \def\url#1{\texttt{#1}}\fi
\expandafter\ifx\csname urlprefix\endcsname\relax\def\urlprefix{URL }\fi
\providecommand{\bibinfo}[2]{#2}
\providecommand{\eprint}[2][]{\url{#2}}

\bibitem{ref:NIST_randomness_tests}
\bibinfo{author}{Rukhin, A.} \emph{et~al.}
\newblock \bibinfo{title}{Nist special publication 800-22: A statistical test
  suite for the validation of random number generators and pseudo random number
  generators for cryptographic applications}.
\newblock \emph{\bibinfo{journal}{NIST Special Publication}}
  \textbf{\bibinfo{volume}{800}}, \bibinfo{pages}{22} (\bibinfo{year}{2010}).

\bibitem{ref:nistSoto}
\bibinfo{author}{Soto, J.} \& \bibinfo{author}{Bassham, L.}
\newblock \bibinfo{title}{Randomness testing of the advanced encryption
  standard finalist candidates}.
\newblock \bibinfo{type}{Tech. Rep.}, \bibinfo{institution}{Booz-Allen And
  Hamilton Inc Mclean Va} (\bibinfo{year}{2000}).

\bibitem{ref:Doganaksoy}
\bibinfo{author}{Do{\u g}anaksoy, A.}, \bibinfo{author}{Sulak, F.},
  \bibinfo{author}{U{\u g}uz, M.}, \bibinfo{author}{{\c S}eker, O.} \&
  \bibinfo{author}{Akcengiz, Z.}
\newblock \bibinfo{title}{New statistical randomness tests based on length of
  runs}.
\newblock \emph{\bibinfo{journal}{{Math. Prob. Engin.}}}
  \textbf{\bibinfo{volume}{2015}}, \bibinfo{pages}{626408}
  (\bibinfo{year}{2015}).

\bibitem{schrijencreating}
\bibinfo{author}{Schrijen, G.-J.} \& \bibinfo{author}{Maes, R.}
\newblock \bibinfo{title}{Creating an efficient random number generator using
  standard sram} .
\end{thebibliography}
\end{document}